# Plasma Medicine and Osmosis

M. N. Shneider[1], M. Pekker[2]

[1]Department of Mechanical and Aerospace Engineering, Princeton University, Princeton NJ 08544
[2]The George Washington University, Northwest Washington, DC 20052

Email:  m.n.shneider@gmail.com

**Abstract**
In this paper, attention is drawn to the importance of accounting for osmotic pressure when analyzing physiological effects on cellular structures in plasma medicine. Interaction of a weakly ionized plasma jet with a saline solution leads to a detectable changes in the saline's ion-molecular composition and hence changes in the osmotic pressure. This, in turn, leads to a stretching or compression of the membrane, depending on the difference of total external and internal pressures. The selective effect of plasma on cells, observed in experiments, is associated with the change in the mechanical properties of membranes (and thereby, a weakening of their protective properties). Corresponding estimates are given in the article.

**Introduction**

In recent decades, plasma medicine has developed significantly. In a number of experimental studies, it has been convincingly shown that the action of a weakly ionized nonequilibrium plasma jet on a physiological solution can selectively affect the cells present within it [1-9]. Special attention should be paid to works in which selective stimulation of apoptosis in cancer cells was observed [4-9].  However, the mechanism by which plasma affects cells in saline is still unclear (see, for example, [1-3]). When considering the interaction of weakly-ionized plasma jets with cellular structures, various physical and chemical processes caused by currents induced in the physiological solution and cells must be taken into account (see, for example, [10]). A well-established experimental fact is that when plasma is exposed to a saline solution, long-lived (tens of minutes or even hours) solvated (hydrated) ions and electrons appear in the solution [11].

In this paper, attention is drawn to the fact that, since cell membranes are permeable to water molecules and poorly permeable to solvated ions, a change in the composition of ions in a physiological solution leads to a change in the osmotic pressure on the cell membrane. This, in turn, leads to a stretching or contraction of the membrane, depending on the ratio of total external and internal pressures.

The effect of the osmotic pressure influence on the vital activity of cells is well known in the physiology of cells [12]. For example, the phenomenon of erythrocyte osmotic fragility (osmotic hemolysis), which consists of the destruction of erythrocyte cells, is relatively well studied. These cells were subjected to osmotic stress by being placed in a hypotonic solution [13, 14]. However, plasma medicine research, as far as we know, has not addressed the role of changing osmotic pressure in the cells due to the saline being exposed to the plasma jet. In this paper, it is shown that the solvated ions, electrons and neutral molecules supplied by the plasma jet may cause a noticeable change in the osmotic pressure and, consequently, a change in the pressure drop across the cell membrane. Since the osmotic pressure depends on the permeability of the membrane to water molecules, then for different membranes, such as those of healthy and diseased cells, osmosis will be different. If the membrane stretching exceeds the critical strength of the membrane, it will break due to unlimited growth of the pores in it [13,14]. The selective effect of plasma on cells (in which only certain types of cells are affected) in a physiological solution is associated with a change in the permeability of the cell membrane when the cell is forcibly stretched or compressed during osmotic pressure changes.

Figure 1 shows a schematic of a typical experimental setup for a plasma jet and cells in a physiological solution in a Petri dish (see, for example, [4-9]). A weakly ionized plasma jet interacting with a physiological solution changes the local ionic and molecular composition of the solution. Plasma-introduced solvated ions and neutral molecules then diffuse into the area where the cells are located. The density of charged particles in a saline solution of electrolyte, corresponding to a living organism, is $n_i \approx 0.3$ M/L $\approx 2 \cdot 10^{26}$ m$^{-3}$ [15]. The direct effect of plasma on cells is insignificant because of the low flux density and the shielding of the electric field by the electrolyte ions. The characteristic Debye screening length, corresponding to a typical ion density $\sim 2 \cdot 10^{26}$ m$^{-3}$, does not exceed 1 nm. However, in principle, it is possible that additional charging of the cell membrane by currents induced in the electrolyte contribute to the critical voltage at which a breakdown occurs in the membrane (its electroporation). This mechanism has been considered in many papers (see, for example, [10, 16-19]); therefore, we will not dwell on it.

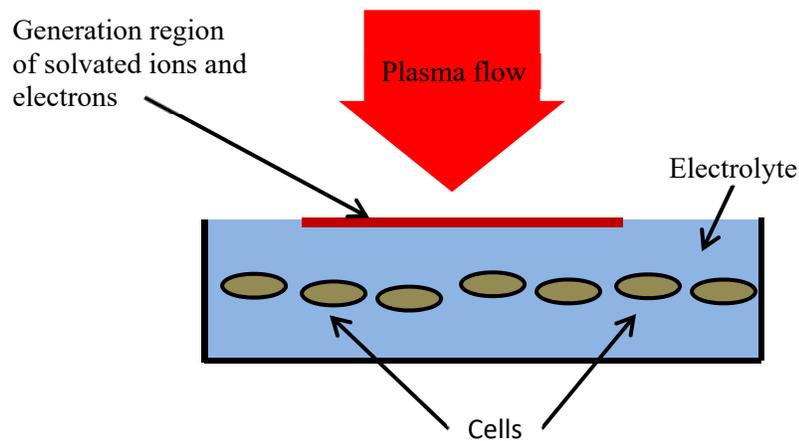

**Fig. 1.** Schematic of the experimental setup for a plasma jet acting on cells in a Petri dish.

Figure 2 shows a simplified scheme of water molecule transport through channels in the cell membrane without and with solvated (hydrated) ions generated by the plasma source. In Fig. 2a, the flow of water molecules from the cell equals the flow of water into the cell. The osmotic pressure drop is zero because the ion densities outside and inside the cell are equal. In Figure 2b, the flow of water molecules from the cell is greater than the flux into the cell and, accordingly, the cell compresses due to mass loss. Compression continues until the flux of water molecules through the membrane equalizes, or in other words, the degree of water salinity inside and outside the cell reaches the same level. The excess amount of ions outside the cell (Figure 2b) leads to a drop in osmotic pressures, which also contributes to cell compression. Reduction of the cell radius due to changing osmotic pressure leads to a decrease in membrane tension. The final part of the article estimates the effect of osmotic pressure on the balance of forces acting on the pores.

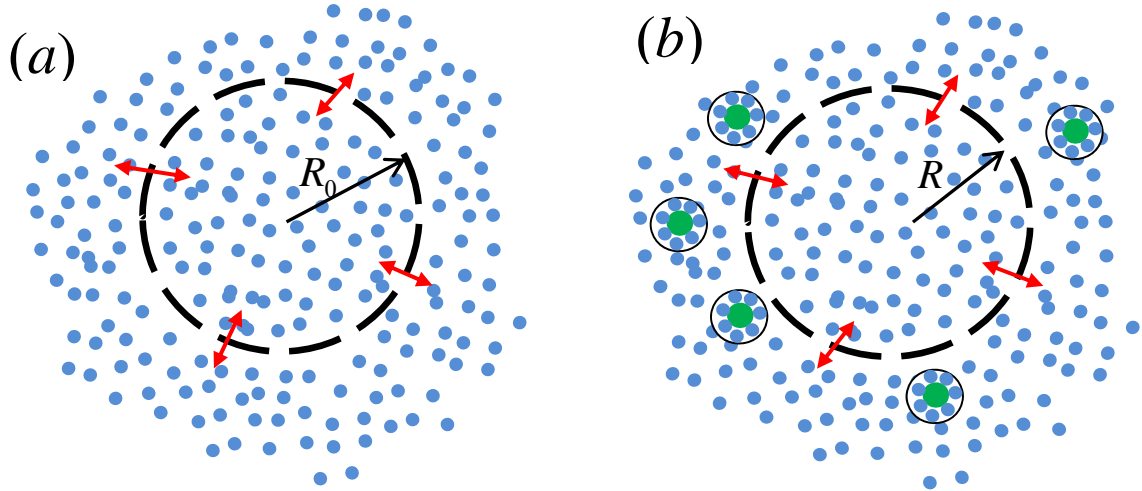

**Fig. 2.** Diffusion of water molecules through the cell membrane. Arrows show the transport of water molecules through the pores in the membrane. (a) - the flow of water molecules into and out of the cell is equal and the osmotic pressure is zero. (b) - the presence of additional solvated (hydrated) ions introduced by the plasma (denoted by smaller circles) reduces the number of free water molecules that can pass through the membrane. The flux of water molecules out of the cell exceeds the flux into the cell. Consequently, the pressure outside the membrane exceeds the inside and osmotic pressure compresses the cell. Compression proceeds until the fluxes through the membrane equalize. Note that the density of "free" water molecules outside and inside the membrane will not be equal, and, as a result, $R_0 > R$.

**Diffusion of Solvated Ions**

Solvated ions and neutral molecules introduced by the plasma source on the surface of the physiological solution (Fig. 1) propagate into the interior of the electrolyte due to diffusion and turbulent mixing. Let us show that, for the characteristic operating time of the plasma source, the solvated ions injected at the surface of the solution are likely to reach the cell surface. For simplicity, consider only diffusion as the slowest process for transporting ions (and neutral molecules) to the surface of cells.

The lifetime of solvated ions and neutral molecules injected into saline is long enough to neglect decay and treat them as stable. For example, the lifetime of ions $NO_3^-$, $NO_2^-$ and molecules $H_2O_2$ is of the order of several hours [20]. Therefore, in the first approximation, we can neglect the processes of ion recombination in a physiological solution. The radius of the region of interaction of the plasma with the surface of the liquid is greater than the distance from the surface to the cells, so the problem of diffusion of salivated charged particles (and neutral molecules) can be described by a one-dimensional diffusion equation:

$$\frac{\partial n}{\partial t} = D \frac{\partial^2 n}{\partial x^2}, \tag{1}$$

with boundary conditions in the injection region

$$D\frac{\partial n}{\partial x}\bigg|_{0, t>0} = J\theta(t_0 - t), \qquad \theta(t_0 - t) = \begin{cases} 1 & at\ t < t_0 \\ 0 & at\ t \geq t_0 \end{cases} \tag{2}$$

and at "infinity"

$$n_\infty = 0. \tag{3}$$

The right-hand side of (2) takes into account the time dependence of the flux of solvated ions generated by the plasma source near the surface of the liquid. It is convenient to introduce dimensionless variables

$$\tau = \frac{t}{t_0}, \quad \xi = \frac{x}{l_0} = \frac{x}{\sqrt{Dt_0}}, \quad \mu = \frac{n}{n_0} = \frac{n}{J}\sqrt{\frac{D}{t_0}}. \tag{4}$$

Equation (1), with boundary conditions (2) and (3), has solution:

$$\mu = \int_0^\tau \frac{1}{\sqrt{\pi(\tau - \zeta)}} \exp\left(-\frac{\xi^2}{4(\tau - \zeta)}\right) \theta(1 - \zeta) d\zeta. \tag{5}$$

Fig. 3 shows the solution (5) for different values of the dimensionless time $\tau$.

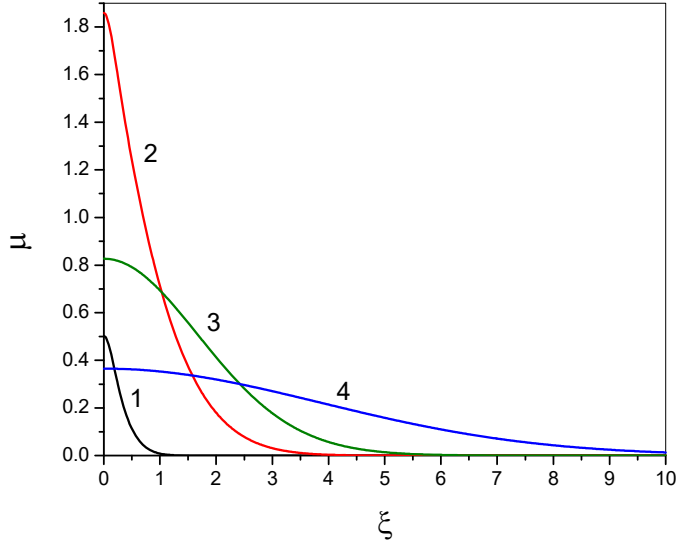

**Fig.3.** Solution of the diffusion equation in dimensionless variables. Line 1 corresponds to $\tau = 0.1$, 2 to $\tau = 1$, 3 to $\tau = 2$, and 4 to $\tau = 4$.

In experiments [4-9] which study the effect of a plasma jet on cells in a Petri dish (Fig. 1), the duration of the plasma jet ranged from 0.5 to 2 minutes, and in [20], up to 60 minutes. For the vast majority of solvated (hydrated ions) in fresh and salt (sea) water at a temperature of ~25 °C, the diffusion coefficient lies within $D \sim 10^{-9} - 5 \cdot 10^{-9}\ m^2/s$ (see, for example, data [21, 22]). Thus, at $t_0 \sim 100$ sec, $l_0 = \sqrt{Dt_0} \sim 0.3-1$ mm, that is, solvated ions can diffuse to the cells that are in the upper layers of physiological solution in a Petri dish (with a typical depth of several millimeters). In fact, the volume containing solvated ions and neutral molecules is much larger because of the deformation of the surface of the liquid by the dynamic pressure of the plasma jet and the resulting hydrodynamic flows.

**The equilibrium radius of the cell**

Suppose, for example, that a cell membrane is permeable only to water molecules, but impermeable to solvated ions. In this case, the equilibrium radius of the cell is determined by the salinity of the solution inside and outside the cell. If the salinity of the solution inside is greater than the salinity outside, the cell absorbs water, and its volume grows until the salinity levels are equal. In the opposite case, the cell expels water, and its volume decreases until the salinity levels are equal.

Let $n_i$ is the sum of the densities of all solvated ions near the cell surface. The relative density of solvated ions (salinity) is

$$\alpha_i = \frac{n_i}{n_w}, \tag{6}$$

where $n_w$ is the local density of water molecules. Let the density of ions near the surface of the cell increase by an amount $\delta n_i$ due to ions injected by the plasma. As a result, the osmotic pressure drop changes and, consequently, the cell radius will decrease from the initial $R_0$ to $R$, at which the water salinity inside and outside the cell and will balance to

$$\alpha_i = \frac{\frac{4}{3}\pi R_0^3 n_i}{\frac{4}{3}\pi R^3 n_w} = \frac{n_i + \delta n_i}{n_w}. \tag{7}$$

Hence, the relative change in the radius of the cell is:

$$\frac{R-R_0}{R_0} = \frac{1}{(1+\delta n_i/n_i)^{1/3}} - 1 \approx -\frac{1}{3}\frac{\delta n_i}{n_i} = -\frac{1}{3}\beta \tag{8}$$

In (8) we took into account that $\delta n_i \ll n_i$, and introduced a dimensionless variable $\beta = \delta n_i / n_i$.

**Estimation of the osmotic pressure drop across the cell membrane**

Let us estimate the pressure drop across the cell membrane resulting from the presence of solvated ions injected by the plasma jet. According to [23], the difference in the osmotic pressure of impurities across the membrane is proportional to the difference in impurity densities:

$$\delta P_i \approx \delta n_i k_B T = \beta n_i k_B T \approx 10^6 \beta \text{ [Pa]}, \tag{9}$$

where $k_B$ is the Boltzmann constant. The quantitative coefficient in (9) was obtained on the assumption that the temperature of the solution $T = 300$.

As shown in [20], the densities of ions $NO_3^-$, $NO_2^-$, and molecules $H_2O_2$, are on the order of 0.1-0.2 mM/L, which corresponds to $\beta \approx (0.7-1.4) \cdot 10^{-3}$, that is, the resultant additional osmotic pressure is 0.7-1.4 kPa. It is known that the critical overpressure at which the phospholipid membrane is destroyed is of the order of 10-12 kPa [24], that is, an order of magnitude greater than our estimate. However, even small changes in pressure can disrupt the mechanical properties of membranes, and therefore, change the transport into or out of the cell. Namely, the changes in the mechanical properties of membranes could be the reason for the selective effect of plasma on cells (for instance, the reason for selective apoptosis) [4-9].

It should be noted that the estimate (9) is valid for weak solutions in which impurities can be considered as an ideal gas that does not interact with the molecules of the solvent. In reality, the ions generated by the plasma interact with the water molecules (are solvated) and therefore, the pressure drop (9) is only a rough estimate.

**Mechanical properties of the membrane and the conditions for growth of pores with regard to osmosis**

One of the channels of cellular metabolism of nutrients within the environment is nanopores, whose size depends on the membrane tension. Deformation of the membrane, expansion or compression, may significantly alter its permeability. Let us consider the contribution of osmotic pressure to the balance of forces that determine the size of the pore. In accordance with [25], the energy of a spherical membrane is:

$$W_K = \frac{1}{2} K_s (A - A_0 - A_p)^2 + 2\pi\gamma R_p, \tag{10}$$

where $A_0 = 4\pi R_0^2$, $A = 4\pi R^2$ are the equilibrium surface area of the spherical unstretched membrane and the surface area of the stretched membrane, respectively; $A_p = \pi R_p^2$ is the area of the round pore of the radius $R_p$; $K_s$ is the stretching modulus, $K_s = 0.2$ J/m$^2$; $\gamma = 10^{-12}$ J/m is the pore line tension. In (10), we need to add a term describing the work that should be done to create the pore. It is obtained from the mechanical work against surface tension forces [16] and electrostatic forces [26]:

$$W_t = \frac{1}{2} K_s \left(4\pi(R^2 - R_0^2) - \pi R_p^2\right)^2 + 2\pi\gamma R_p - \pi R_p^2 \Gamma + \pi R_p^2 h \left(\frac{\varepsilon_w - \varepsilon_m}{\varepsilon_m \varepsilon_w}\right) \frac{\Sigma^2}{\varepsilon_0}, \tag{11}$$

where $\Gamma = 10^{-3}$ N/m is the surface tension at the interface between the non-conducting and conducting liquids [16], $h$ is the half-thickness of membrane, $\Sigma$ is the membrane surface charge, $\varepsilon_0$ is the dielectric permittivity of vacuum; $\varepsilon_m$, $\varepsilon_w$, are the dielectric permittivities in membrane and electrolyte, respectively. The force acting on the pore is:

$$F_t = 16\pi^2 K_s R_0^2 \frac{R-R_0}{R_0} R_p - 2\pi\gamma + 2\pi\Gamma R_p - 2\pi h \left(\frac{\varepsilon_w - \varepsilon_m}{\varepsilon_m \varepsilon_w}\right) \frac{\Sigma^2}{\varepsilon_0} R_p. \quad (12)$$

In (12), we took into account that $|R - R_0| \ll R_0$ and neglected the change in surface area of the membrane due to pore formation. For $R < R_0$, the first term leads to compression of the membrane, whereas for $R > R_0$, the first term leads to expansion.

If we neglect the change in the membrane surface area due to osmotic pressure, then taking into account (8), we obtain:

$$F_t = -\frac{16\pi^2}{3}\beta K_s R_0^2 R_p - 2\pi\gamma + 2\pi\Gamma R_p - 2\pi h \left(\frac{\varepsilon_w - \varepsilon_m}{\varepsilon_m \varepsilon_w}\right) \frac{\Sigma^2}{\varepsilon_0} R_p. \quad (13)$$

Comparing the first and third terms of the right-hand side, we obtain that when $\beta > 10^{-2}$, the force acting on the pore is always negative. Although the value of $\beta$ in the experiments [20] is an order of magnitude smaller than the above value, the density of the solvated ions may be sufficient to be taken into account at other possible plasma parameters.

Let us take into account the effect of the pressure drop $\delta P_i$ (9) on the balance of forces acting on the membrane. The equilibrium radius of the pore is determined by the balance of forces acting on it:

$$-\frac{\partial W_t}{\partial R} + 4\pi R_0^2 \delta P_i = -64\pi^2 K_s R_0^2 (R - R_0) + 4\pi R_0^2 \delta P_i = 0. \quad (14)$$

From (14) we find the relative variation of the radius of the membrane:

$$\frac{(R-R_0)}{R_0} = \frac{\delta P_i}{16\pi K_s R_0} \quad (15)$$

Substituting (15) into (12) we obtain:

$$F_t = \pi \delta P_i R_0 R_p - 2\pi\gamma + 2\pi\Gamma R_p - 2\pi h \left(\frac{\varepsilon_w - \varepsilon_m}{\varepsilon_m \varepsilon_w}\right) \frac{\Sigma^2}{\varepsilon_0} R_p. \quad (16)$$

Since the osmotic pressure compresses the cell, i.e. $\delta P_i < 0$, we obtain that at $|\delta P_i| > \frac{2\Gamma}{R_0} \approx \frac{2 \cdot 10^{-3}}{R_0}$, $F_t$ is negative, at any possible pore radius. In other words, the pores cannot grow.

**Conclusions**

We have shown that the selective effect of a plasma jet on living cells in a physiological solution can be related to a change in the osmotic pressure difference across the cell membrane, as a result of the injection of additional long-lived solvated (hydrated) ions by the plasma. These ions result in an outflow of water from the cell volume, a decrease in its radius, and additional pressure on the membrane. These factors change the balance of forces acting on the membrane and the development of pores in it, which ultimately affects the viability of the cell.